\documentclass{emulateapj}
\usepackage[normalem]{ulem} 
\usepackage{soul} 
\usepackage{natbib}
\usepackage{apjfonts}

\usepackage{amsmath}

\usepackage{epsfig}
\usepackage{graphicx}
\usepackage{amssymb}
\usepackage{aas_macros}
\usepackage{color}

\newcommand{\kms}{\,{\rm km \, s^{-1}}}
\newcommand{\oversim}[2]{\protect{\mbox{\lower0.5ex\vbox{%
   \baselineskip=0pt\lineskip=0.2ex
   \ialign{$\mathsurround=0pt #1\hfil##\hfil$\crcr#2\crcr\sim\crcr}}}}} 
\newcommand{\simgreat}{\mbox{$\,\mathrel{\mathpalette\oversim>}\,$}} 
\newcommand{\simless} {\mbox{$\,\mathrel{\mathpalette\oversim<}\,$}} 

\catcode`\"=\active\let"=\"  

\def\3{{\ss} }

\def\c12{{1\over 2}}

\def\d{{\rm d}}   
   
\def\plusplus{\raise 0.3ex\hbox{${\scriptstyle ++}$}{}}   
   
\def\ve{\varepsilon}
   
\begin{document}   
   
\title{A statistical method for measuring the Galactic potential and testing gravity with cold tidal streams}   
   
\author{Jorge Pe\~{n}arrubia\altaffilmark{1,2}, Sergey E. Koposov\altaffilmark{2,3} \& Matthew G. Walker\altaffilmark{4}}

\email{jorpega@iaa.es}
\altaffiltext{1}{Ram\'on y Cajal Fellow, Instituto de Astrof\'isica de Andalucia-CSIC, Glorieta de la Astronom\'ia s/n, 18008, Granada, Spain}
\altaffiltext{2}{Institute of Astronomy, University of Cambridge, Madingley Road, Cambridge, CB3 0HA, UK}
\altaffiltext{3}{Sternberg Astronomical Institute, Moscow State University, Universitetskiy pr. 13, 119992 Moscow, Russia}
\altaffiltext{4}{Hubble Fellow, Harvard-Smithsonian Center for Astrophysics, 60 Garden St., Cambridge, MA 02138, USA}


\begin{abstract}   
We introduce the Minimum Entropy Method, a simple statistical technique for
constraining the Milky Way gravitational potential and simultaneously testing
different gravity theories directly from 6D phase-space surveys and without
adopting dynamical models. 
We demonstrate that orbital energy distributions that are separable (i.e. independent of position) have an associated entropy that increases under wrong assumptions about the gravitational potential and/or gravity theory.
Of known objects, `cold' tidal streams from low-mass progenitors follow orbital distributions that most nearly satisfy the condition of separability. Although the orbits of tidally stripped stars are perturbed by the progenitor's self-gravity, systematic variations of the energy distribution can be quantified in terms of the cross-entropy of individual tails, giving further sensitivity to theoretical biases in the host potential.
The feasibility of using the Minimum Entropy Method to test a wide range of
gravity theories is illustrated by evolving restricted N-body models in a
Newtonian potential and examining the changes in entropy introduced by Dirac,
MONDian and $f(R)$ gravity modifications. 
\end{abstract}   

\section{Introduction} \label{sec:intro}
Studies of the Milky Way potential usually involve the construction of dynamical models under a given law of gravity (e.g. Newton's). Knowledge about the amount and distribution of mass in our Galaxy is typically gained by solving the equations of motion in a given gravitational potential and contrasting the body motions predicted by the theoretical models against the observed kinematics of different sorts of tracers (i.e. bulk gas motions; stellar radial velocities and proper motions).


Statistical tools provide an alternative methodology for inferring the dynamical properties of gravitational systems directly from observational data sets and without an {\it a priori} understanding of the motions of kinematic tracers in the system that is being observed. 
In this contribution it will be shown that under some special conditions it is possible to derive statistical inferences about the Milky Way potential without solving the equations that govern the motion of stars in the Galaxy, but by merely assuming that the orbital energy is an integral of motion. Gravity tests can thus be incorporated in the analysis in a quite straightforward manner.

 The method of statistical inference outlined here focuses on stellar tidal debris, i.e. stars that are tidally stripped from gravitationally-bound objects. These stars have a unique dynamical property: although they sample very large volumes of the host's phase-space (position-velocity), their orbits are close to that of the progenitor system. The Sagittarius dwarf galaxy provides a dramatic example, as it is currently shedding stars to the Milky Way halo in the form of a tidal stream that completely wraps around our Galaxy (e.g. Koposov et al. 2011b and references therein). It has been extensively shown that the location and kinematics of tidal tails can be used to put strong constraints on the host potential under the assumption that tidal tails follow single orbits (e.g. Law et al. 2005, 2009; Pe\~narrubia et al. 2005; Jin \& Lynden-Bell 2007; Binney 2008; Eyre \& Binney 2009; Koposov et al. 2010). 

Unfortunately, most tidal debris will be too faint to be detected as coherent over-densities (e.g. tidal tails, clouds or shells) in photometric or spectroscopic surveys, either because their progenitors contained a small number of stars, or because the stars were stripped a long time ago. Indeed, the dynamical evolution of tidal debris adds considerable complexity to the detection, follow-up and subsequent dynamical analysis of these systems, for tidally-stripped stars tend to progressively fill the allowed phase-space volume of the progenitor's orbit, becoming in the process kinematically colder and spatially more sparse (e.g. Helmi \& White 1999). 

In contrast, the evolution of tidal debris in the space of integrals of motion is remarkably quiescent. In this space tidally-stripped stars are distributed tightly about the integrals of motion of the progenitor object regardless of how far in the past the disruption event occurred; a property that holds even in 
potentials that vary slowly with time (Pe\~narrubia et al. 2006). 

Some integrals of motion, like the angular momentum in a spherical potential, are quantities that can be measured directly from phase-space coordinates. Others, like the orbital energy in a static potential, require additional hypothesis of the behaviour of gravity in the Milky Way. Recently, Binney (2008) found that wrong theoretical assumptions on the Milky Way potential increase the scatter in the orbital energies inferred from the locations and velocities of individual stars along tidal tails\footnote{McMillan \& Binney (2008) show that a similar principle applies in the action-angle space, where wrong assumptions on the local potential alter the characteristic periodicity (or ``beats'') of the patterns shown by tidal debris as they cross the solar neighbourhood.}. This statistical property enables him to construct an algorithm for measuring the Galactic potential by minimizing the r.m.s. energies derived from fitting single orbits to tidal tails. However, this algorithm relies on the assumption that all stars on the tails have the same orbital energy in the true gravitational potential.

 Here it is demonstrated that Binney's (2008) results can be extended to tidal debris with arbitrary energy distributions as long as these distributions are separable (i.e. invariant) in space. 
This assumption may be accurate for stellar streams whose progenitors have low dynamical masses, the so-called 'cold' tidal streams (Pe\~narrubia et al. 2006; Eyre \& Binney 2009). Also, it will be shown that these systems can be used to measure the Milky Way gravitational potential as well as to test different gravity theories simultaneously, and without having to integrate the equations of motion. Indeed, our method provides a simple tool to constrain the Milky Way gravity through a direct inspection of the phase-space coordinates of cold tidal streams, thereby avoiding the daunting task of solving for the dynamics of stars in host galaxies with a generic mass distribution under a suite of gravity theories.

Dropping the necessity to fit orbits to the locations and velocities of streams brings a further advantage, namely the fact that our method can be equally applied to phase-space substructures that can be identified as tidal tails, as well as to substructures that have 'dissolved' in the Galactic halo, but which may be detected as substructures in the space of integrals of motion, as long as they originate from low-mass progenitors.

Before introducing our method we wish to remark that the technique outlined in this contribution requires full 6D phase-space information. At present most stars with measured position-velocity vectors are located in the solar neighbourhood, which severely limits its applicability. However, this unfortunate situation will likely improve in the near future thanks to the advent of Gaia, an European mission that will provide astrometric data for more than one billion stars with unprecedented accuracy (Perryman et al. 2001). Gaia data in combination with complementary spectroscopic surveys (e.g. Gaia-ESO) is expected to reveal a large number of substructures associated with past accretion events (Brown et al. 2005; Mateu et al. 2011). 

The paper arrangement is as follows: we develop the core idea of our method in \S\ref{sec:entropy}. Tests are shown in \S\ref{sec:numtest}. We go through a brief discussion about applications and limitations in \S\ref{sec:self} and~\S\ref{sec:disc}.

\section{The entropy of tidal debris}\label{sec:entropy}
In statistical mechanics, entropy measures the degree to which the probability of the system is spread out over different possible states. The concept of entropy has been widely used in Physics (e.g. Wehrl 1978; Jaynes 1957) to determine the behaviour of macroscopic systems in equilibrium, or close to equilibrium. For example, galaxies undergoing phase-space mixing tend to evolve toward an energy configuration that maximizes entropy\footnote{More specifically, Tremaine, Hen\'on \& Lynden-Bell (1986) show that the equilibrium configuration is that which maximizes {\it all} H-functions $H\equiv -\int C(F)\d{\bf r}\d{\bf v}$, where $F({\bf r},{\bf v})$ is the coarse-grained distribution function, and $C(F)$ is any convex function with $C(0)=0$. Entropy corresponds to the case $C(F)=F\ln F$}.
 
In this work we are interested in stellar systems that follow a narrow energy distribution and have, by construction, low entropy. Stars that are tidally stripped from gravitationally-bound, low-mass objects and orbit about the host galaxy potential provide an example of systems with low entropy. Given that biases in the theoretical modelling of our Galaxy tend to {\it increase} the range of energies sampled by tidal debris (Binney 2008), it appears therefore natural to use entropy as statistically-meaningful quantity to identify those biases. In addition, entropy has special properties that will help us below to tackle difficult aspects of the study. For example, its additivity, i.e the fact that the combined entropy of two independent systems equals the sum of the individual entropies (see \S II.E of Wehrl 1978), helps to combine information on the Milky Way potential gathered from individual debris systems (\S\ref{sec:dh_mult}), as well as to analyze the effects of the progenitor's self-gravity (\S\ref{sec:self}).

It is illustrative to re-interpret Binney (2008)'s results in terms of entropy. Consider a suite of $N_\star$ stars orbiting in a potential $\Phi({\bf r})$ with phase-space coordinates $\{{\bf r}_i,{\bf v}_i\}_{i=1}^{N_\star}$ and an orbital energy per unit mass $E_i=E_0$ for all $i$, where 
\begin{eqnarray}
E_i=\frac{{\bf v}_i^2}{2} + \Phi({\bf r}_i).
\label{eq:ener}
\end{eqnarray}
Given that all stars have the same energy in the true potential, the energy distribution follows a Dirac's delta function per construction. Binney (2008) shows that wrong assumptions about $\Phi({\bf r})$ must necessarily yield a scattered sample of energies unless all the stars are located at the same position. The larger the departure from the true underlying potential, the higher the rms variation in energy. 
Using entropy one would conclude that, since by definition a delta function has an entropy of minus infinity, the entropy of the {\it biased} energy distribution can only increase owing to poor choices in the potential model. Clearly, the stronger the bias the larger the scatter, and thus the higher the measured entropy must be. 

In \S\ref{sec:dh_single} we extend this principle to stellar systems whose energy distribution $f(E)$ is differentiable and separable in energy and space\footnote{Note that a similar analysis can be developed in the space of actions.}. Mathematically this condition implies that the probability that a star having an energy $E$ at a location ${\bf r}$ can be written as $p(E,{\bf r})=f(E)g({\bf r})$, where $g({\bf r})$ is the probability that a star has position ${\bf r}$. Both probability functions are normalized so that $\int f(E)\d E = \int g({\bf r})\d^3r=1$. It is convenient to define the relative potential as $\Psi=-\Phi+\Phi_\infty$, and the relative energy as $\varepsilon=-E + \Phi_\infty$. Here $\Phi_\infty$ is an arbitrary constant that sets the boundary conditions for the energy distribution $f(\ve)$. 

The fact that the Milky Way potential is unknown introduces a {\it model-dependent bias} ($\delta \Phi$) in the orbital energy derived from phase-space measurements
\begin{equation}
\tilde \ve({\bf r}) = \ve({\bf r}) +\delta \Phi({\bf r}).
\label{eq:phibias}
\end{equation}
In what follows we use tildes to denote measured quantities, whereas symbols without tildes denote true (i.e. unbiased) quantities.

\subsection{Single component}\label{sec:dh_single}
For systems with separable energy distributions, the biased energy distribution relates to the true one as $\tilde p(\ve,{\bf r})=p[\ve - \delta \Phi({\bf r}),{\bf r}]=f[\ve - \delta \Phi({\bf r})]g({\bf r})$. Let us expand the measured energy distribution $\tilde f(\ve)$ at order $\mathcal{O}[(\delta \Phi)^2]$

\begin{eqnarray}
\label{eq:fbias}
\tilde f(\ve) = \int f[\ve - \delta \Phi({\bf r})]g({\bf r})\d^3{\bf r}\approx \\ \nonumber
 f(\ve)\int \bigg[1 - \delta \Phi({\bf r}) \frac{f'(\ve)}{f(\ve)} + \frac{\delta \Phi^2({\bf r})}{2}\frac{f''(\ve)}{f(\ve)}\bigg]g({\bf r})\d^3{\bf r}=\\ \nonumber
f(\ve)\bigg[1 - \langle \delta \Phi \rangle\frac{f'(\ve)}{f(\ve)} + \frac{\langle \delta \Phi^2\rangle}{2}\frac{f''(\ve)}{f(\ve)}\bigg].
\end{eqnarray}
Here brackets denote volume-averaged quantities, i.e. $\langle x \rangle = \int x g({\bf r})\d^3r$, whereas $f'\equiv \d f/ \d\ve$ and  $f''\equiv \d^2 f/ \d\ve^2$. 

The entropy of the measured energy distribution is by definition
\begin{eqnarray}
\tilde H = - \int \d\ve \tilde f(\ve) \ln[\tilde f(\ve)].
\label{eq:hbias}
\end{eqnarray}

After approximating $\ln (1+x)\approx x-x^2/2$ for $x\ll 1$ and some algebra, the combination of eq.~(\ref{eq:fbias}) and~(\ref{eq:hbias}) at $O[(\delta \Phi)^2]$ results in
\begin{eqnarray}
\label{eq:hbias_ext}
\tilde H \simeq H + \langle \delta \Phi\rangle \int \d\ve f'(\ve)[1+\ln f(\ve)] \\ \nonumber
-\frac{\langle \delta \Phi\rangle^2}{2}\int \d\ve  f(\ve)\bigg[\frac{f'(\ve)}{f(\ve)}\bigg]^2 -
\frac{\langle \delta \Phi^2\rangle}{2} \int \d\ve f''(\ve)[1+\ln f(\ve)]   .
\end{eqnarray}

The integrals including the term $(1+\ln f)$ can be easily solved by parts. 
Adopting boundary conditions so that $\displaystyle\lim_{\Phi\rightarrow 0} f\ln f = \lim_{\Phi\rightarrow \Phi_\infty} f\ln f =\lim_{\Phi\rightarrow 0}f'\ln f = \lim_{\Phi\rightarrow \Phi_\infty} f'\ln f =0$ one can readily show that the first term is zero
\begin{eqnarray}
\label{eq:kk1}
\int \d\ve f' (1+\ln f) = \big(f\ln f\big)_0^{\Phi_\infty}  = 0, \nonumber
\end{eqnarray}
whereas
\begin{eqnarray}
\label{eq:kk2}
\int \d\ve f'' (1+\ln f) = \big[f'(1+\ln f)\big]_0^{\Phi_\infty}- \int \d\ve f\bigg[\frac{f'}{f}\bigg]^2 = \nonumber \\ 
 - \int \d\ve f\bigg[\frac{f'}{f}\bigg]^2. \nonumber
\end{eqnarray}

Hence eq.~(\ref{eq:hbias_ext}) becomes
\begin{eqnarray}
\tilde H \simeq H + \frac{\langle \delta \Phi^2\rangle - \langle \delta \Phi\rangle^2}{2}\int \d\ve f(\ve)\bigg[\frac{f'(\ve)}{f(\ve)}\bigg]^2 \equiv H + \frac{\sigma_\Phi^2}{2 \sigma_\ve^2},
\label{eq:deltah}
\end{eqnarray}
where the quantity $\sigma^2_\Phi\equiv \langle \delta \Phi^2\rangle - \langle \delta \Phi\rangle^2$ is the dispersion in the energy distribution that arises from the bias in our Galaxy potential (eq.~\ref{eq:phibias}) over the volume occupied by the tidal debris sample, and $\sigma_\ve^{-2}\equiv \int \d\ve f(\ve)[f'(\ve)/f(\ve)]^2$ is related to the internal dispersion of the energy distribution in the unbiased Galaxy potential. For example, for a Gaussian distribution $f(\ve)=1/\sqrt{2\pi \sigma^2}\exp[-\ve^2/(2\sigma^2)]$ the correspondence is direct, as $\sigma_\ve^{-2}=\int \d\ve f(\ve)[f'(\ve)/f(\ve)]^2 = \sigma^{-2}$. 

Eq.~(\ref{eq:deltah}) shows a few interesting points. First, since both quantities $\sigma_\Phi^2$ and $\sigma_\ve^2$ are positive, we find that {\it any bias introduced in our Galaxy model yields an increase in the entropy of the energy distribution}.
Second, the choice of $\Phi_\infty={\rm const}$ does not alter the value of the entropy, as adding a constant bias to the potential leaves the dispersion of orbital energies unchanged.  
 Third, because $\Delta H\simeq  1/2 (\sigma_\Phi/\sigma_\ve)^2$ we find that the change in entropy is fairly sensitive to biases in the potential averaged over the region probed by the tracer population. 
This is a remarkable property of the entropy, suggesting that {\it minimization of entropy for stellar systems with separable energy distributions provides a powerful method to measure the Milky Way potential}. 
 Cold tidal debris from low-mass progenitors may be an example of such systems because they occupy a reduced volume in the integral-of-motion space, but they sample large volumes of phase-space (see \S\ref{sec:self}). 
\begin{figure*}
\includegraphics[width=174mm, height=174mm]{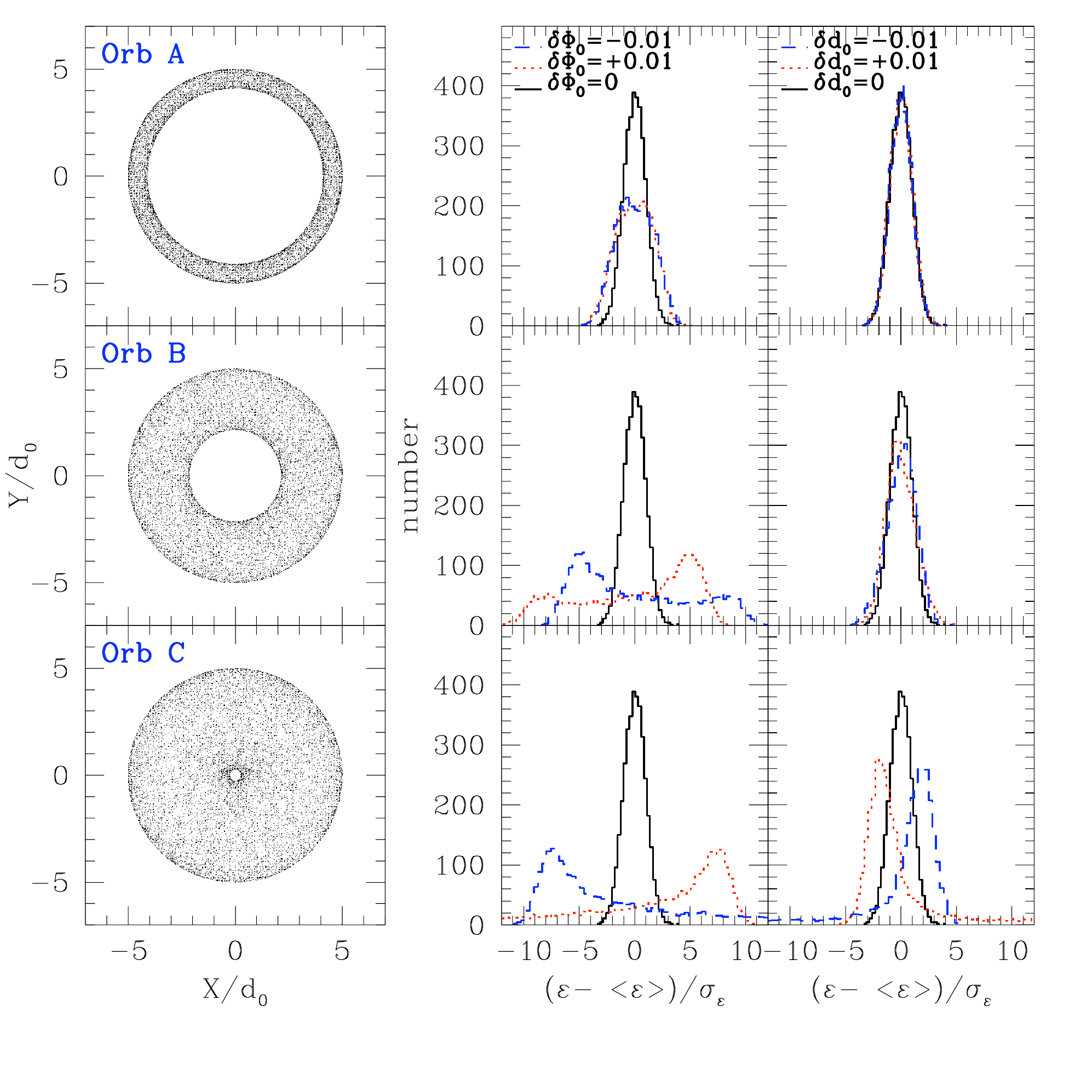}
\caption{{\bf Left column}: Spatial distribution of particles that follow Gaussian orbital energy distribution with a dispersion $\sigma_\varepsilon=10^{-3}$ for orbits with three different orbital pericentres (rows). The orbital apocentre of these models is $5d_0$. {\bf Middle column}: Effects on the orbital energy distribution of introducing a bias in the value of $\Phi_0$. The un-biased distribution is Gaussian (solid lines), whereas dotted and dashed lines denote biases of $+1\%$ and $-1\%$ in the value of $\Phi_0$, respectively. Note that the changes in the measured energy distribution become more prominent as the orbital eccentricity increases.  {\bf Right column}: Effects of a bias in the value of $d_0$. As expected, the changes in the energy distribution are small in models where all particles are located at $r\gg d_0$.}
\label{fig:xyz}
\end{figure*}
\subsection{Multiple components}\label{sec:dh_mult}
In practice it may prove challenging to isolate the energy distribution of a single population of stellar debris from the Milky Way background, or from a population of distinct structures overlapping in a given energy range. It is thus instructive to repeat the above calculations for an energy distribution of the form $f(\ve)=\alpha f_1(\ve) + (1-\alpha)f_2(\ve)$, where $0\le \alpha\le 1$. The measured distribution is thus $f(\ve,{\bf r}) = \alpha f_1(\ve)g_1({\bf r}) + (1-\alpha)f_2(\ve)g_2({\bf r})$.

After some algebra it is straightforward to show that a bias in the galaxy potential of the form given in eq.~(\ref{eq:phibias}) leads to a change in the entropy
\begin{eqnarray}
\label{eq:deltahmult}
\Delta H \simeq \alpha\langle \delta \Phi_1\rangle \int \d\ve f'_1[1+\ln f] 
-\alpha\frac{\langle \delta \Phi_1^2\rangle}{2}\int \d\ve f_1''[1+\ln f] \nonumber \\
- \alpha^2\frac{\langle \delta \Phi_1\rangle^2}{2}\int \d\ve f\bigg(\frac{f_1'}{f}\bigg)^2 
+ (1-\alpha)\langle \delta \Phi_2\rangle \int \d\ve f'_2[1+\ln f] \nonumber \\
-(1-\alpha)\frac{\langle \delta \Phi_2^2\rangle}{2}\int \d\ve f_2''[1+\ln f] - (1-\alpha)^2\frac{\langle \delta \Phi_2\rangle^2}{2}\int \d\ve f\bigg(\frac{f_2'}{f}\bigg)^2 \nonumber \\
-\alpha(1-\alpha)\langle \delta \Phi_1\rangle\langle \delta \Phi_2\rangle\int \d\ve \frac{f_1' f_2'}{f}.
\end{eqnarray}

Note that now the terms that contain the linear variations of the potential, $\langle \delta \Phi\rangle$, in general do not vanish. Hence the entropy of the function $\tilde f(\ve)$ may increase or decrease depending on the potential bias and our choice of $\Phi_\infty$.  These terms only vanish if (i) the structures defined by $f_1(\ve)$ and $f_2(\ve)$ follow the same spatial distribution, i.e. $g_1({\bf r})=g_2({\bf r})$ so that $\langle \delta \Phi_1\rangle= \langle \delta \Phi_2\rangle$ and $\langle \delta \Phi_1^2\rangle= \langle \delta \Phi_2^2\rangle$; or (ii) if the energy distributions of the substructures do not overlap, so that $\int \d\ve f'_i[1+\ln f] = \int_{\Delta \ve_i} \d\ve f'_i[1+\ln f_i]$; and $\int \d\ve f(f_i'/f)^2 = \int_{\Delta \ve_i}\d\ve f_i(f_i'/f_i)^2$, where $\Delta \ve_i$ is the range of energies occupied by the $i_{\rm th}$ substructure. In this case the crossed terms vanish, $\int \d\ve (f'_i f'_j)/f =0$, and by analogy with \S\ref{sec:dh_single} we find the entropy variation of $N_s$ {\it non-overlapping} structures is $\Delta H\approx 1/2 \sum_{i=1}^{N_s} \alpha_{i} (\sigma_{\Phi,i}/\sigma_{\ve,i})^2$, where $\sum_{i=1}^{N_s}\alpha_i=1$. This is indeed an interesting result, for it allows us to combine in a simple way the constraints on the Milky Way potential derived from {\it distinct} substructures under the only condition that their energy distributions do not overlap with each other. 

Finally, eq.~(\ref{eq:deltahmult}) shows that, as in many other problems in Physics, it will be necessary to clean out the sample of stellar tidal debris from Milky Way background objects, as that is a case of two overlapping distributions by definition. To this end it may help to consider additional integrals of motion (e.g. the vertical component of the angular momentum if the potential is axi-symmetric), and/or metal compositions. Also, at some point it may be worth extending our analysis to the space of actions (e.g. McMillan \& Binney 2008), where the background subtraction may be more straightforward.

\section{Tests}\label{sec:numtest}
In this Section we devise a number of numerical experiments that aim to test the analytical results enclosed in \S\ref{sec:entropy}. For simplicity we adopt an unbiased energy distribution that is Gaussian, $f(\varepsilon)=1/\sqrt{2\pi\sigma_\varepsilon^2}\exp[-(\varepsilon-\varepsilon_{\rm orb})^2/(2 \sigma_\varepsilon^2)]$, where $\varepsilon_{\rm orb}$ is the orbital energy of the disrupted system and $\sigma_\varepsilon$ its dispersion. Recall, however, that the results obtained in \S\ref{sec:entropy} apply to {\it any} energy distribution that is differentiable in the range $[0,\Phi_\infty]$ and separable in energy and space.

\subsection{Set-up}
Again for simplicity we adopt a host galaxy potential that is spherical and does not vary with time
\begin{eqnarray}
\Phi(r) =\Phi_0\ln(d_0^2 + r^2).
\label{eq:poten}
\end{eqnarray}

Subsequently, we construct suites of $10^4$ particles initially placed at an initial radius $r_0$. The initial velocities $v=\sqrt{2[-\Phi(r_0) - \varepsilon]}$ are chosen so that the energy distribution is Gaussian. The mean of the energy distribution is $\varepsilon_{\rm orb}=-v_0^2/2-\Phi(r_0)$, where $v_0=\xi v_c=\xi \sqrt{r d\Phi/dr}$, and $v_c$ is the circular velocity of the host. The parameter $\xi\le 1$, so that all our orbits are initially at apocentre. 

The initial azimuthal location of the particles is random. The integration time falls between 1 and 10 $t_{\rm cr}$, where $t_{\rm cr}=r_0/v_0$. The equations of motion are solved using a leap-frog scheme with a time-step that is chosen to provide an energy conservation $\Delta \varepsilon \simless 10^{-1}\sigma_\varepsilon$.

Finally, throughout this Section we adopt the N-body units $G=\Phi_0=d_0=1$.

\begin{figure}
\includegraphics[width=84mm]{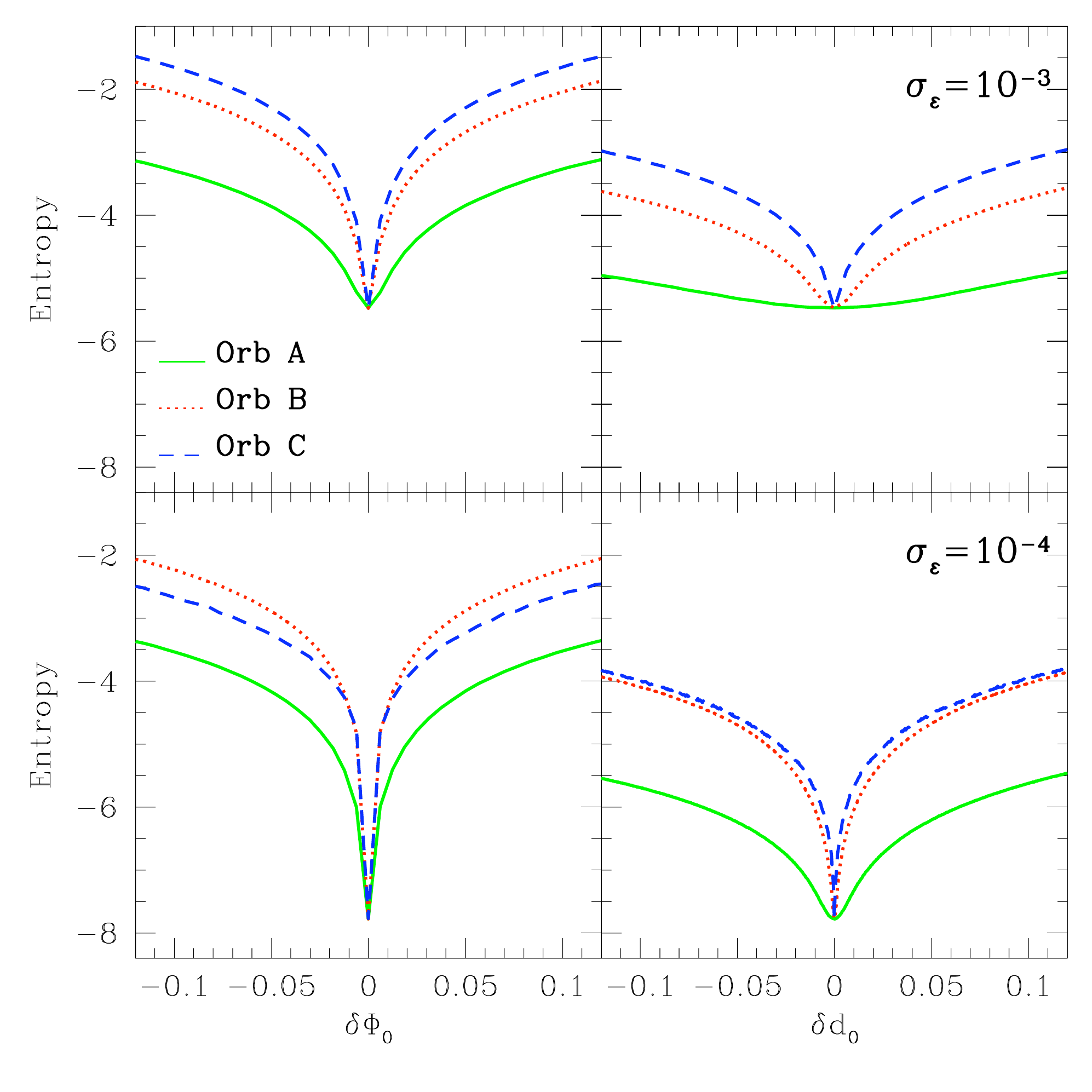}
\caption{Entropy of the orbital energy distribution as a function of bias in the value of $\Phi_0$ (left column; $\delta d_0=0$) and $d_0$ (right column; $\delta \Phi_0=0$) for three different orbits. The entropy of the unbiased ($\delta \Phi_0=\delta d_0=0$) Gaussian distribution is $H=1/2[\ln(2\pi \sigma_\varepsilon^2) + 1] \simeq -5.49$ for $ \sigma_\varepsilon=10^{-3}$, and $H\simeq -7.79$ for  $ \sigma_\varepsilon=10^{-4}$. Note the strong sensitivity of the entropy to biases in the potential parameters as $\delta \Phi_0$ and $\delta d_0$ approach zero (i.e. unbiased potential).  }
\label{fig:ent_4}
\end{figure}

\subsection{Models}\label{sec:models}
For illustrative purposes all the orbits considered here have an apocentre $r_0=5d_0=5$ (we note, however, that a different choice of $r_0$ would not change the qualitative results shown below). By varying the parameter $\xi$ one can change the orbital eccentricity of our models in a simple way. Values of $\xi=0$ and 1 yield radial and circular orbits, respectively. To illustrate the method outlined in \S\ref{sec:entropy} we consider models with $\xi=0.1, 0.6$ and 0.9, which respectively yield orbital pericentres~0.27 (orbit C), 2.14 (orbit B) and 4.11 (orbit A). Their mean orbital energy is $\varepsilon_{\rm orb}\simeq -4.11, -3.60$ and $-3.27$.

We choose energy dispersion values for our models that are are representative of those expected for the tidal debris of dwarf spheroidal galaxies. 
The dispersion of the orbital energy distribution relative to the Galaxy potential roughly scales as $\sigma_\varepsilon\sim \sigma_v^2/v_{\rm max}^2$, where $\sigma_v$ is the original velocity dispersion of the disrupted system and $v_{\rm max}$ is the peak velocity of the Milky Way rotation curve. Adopting $v_{\rm max}=220\kms$ we expect $\sigma_\varepsilon\sim 10^{-4}$--$10^{-3}$ for dwarf spheroidal galaxies, which show velocity dispersions in the range $\sim 3$ and 10$\kms$, respectively (e.g. Mateo 1998; Simon \& Geha 2007; Koposov et al. 2011a). Although low-mass globular clusters can have velocity dispersions as low as $\sigma_v\sim 1\kms$, which correspond to $\sigma_\varepsilon\sim 10^{-5}$, here we only consider intermediate values for the energy dispersion, $\sigma_\varepsilon=10^{-4}$ and $10^{-3}$.

\subsection{Biases arising from the potential parameters}\label{sec:param}
The spatial distribution of test particles for the three orbits considered here is plotted in the left column of Fig.~\ref{fig:xyz}. As shown by the solid lines in the middle and right columns, these particles follow a Gaussian energy distribution in the unbiased ($\delta \Phi_0=\delta d_0=0$) potential. The middle column illustrates the effect of introducing a small bias in the value of $\Phi_0$ on the shape of the energy distribution. Note that biases in the potential change both the mean orbital energy $\langle \varepsilon \rangle$ and the energy dispersion of the particles. However, by plotting the energy distribution as a function of $(\varepsilon - \langle \varepsilon \rangle)$ we can only appreciate variations in the latter. Red-dotted and blue-dashed lines correspond to values of $\tilde \Phi_0=1.01\Phi_0$ and $\tilde \Phi_0=0.99\Phi_0$, respectively. As one would expect, the shape of the energy distribution changes more strongly as the orbital eccentricity increases, and thus the range of orbital radii, increase. The right column shows the effects of changing $d_0$. Here it is interesting to observe that the energy distribution barely changes if all particles move on orbits with $r\gg d_0$ (orbit A), highlighting the fact that constraints on a given potential parameter can only be derived from orbits that are sensitive to variations in that particular parameter. Indeed, as more particles are located at $r\sim d_0$ (e.g. orbit C) changes in $\tilde f(\varepsilon)$ become again visible.

The entropy associated to the orbital energy distribution of these models (eq.~\ref{eq:hbias}) is calculated using a simple trapezoidal rule. The results are plotted as a function of biases in $\Phi_0$ and $d_0$ in the left and right columns of Fig.~\ref{fig:ent_4}, respectively. Rows correspond to models with an energy dispersion $\sigma_\varepsilon=10^{-4}$ and $10^{-3}$.  
This Figure illustrates two interesting points. First, the entropy is indeed minimum for the unbiased potential, as analytically predicted in \S\ref{sec:entropy}. Biases in any of the potential parameters translate into an increase of entropy that is more marked for models with intrinsically narrow energy distributions. Second, the increase in entropy becomes more pronounced as $\delta \Phi_0$ and $\delta d_0$ approach zero. This is an important result, as it suggests that entropy minimization may provide a powerful method to measure the gravitational potential of galaxies with high accuracy.

\begin{figure}
\includegraphics[width=84mm]{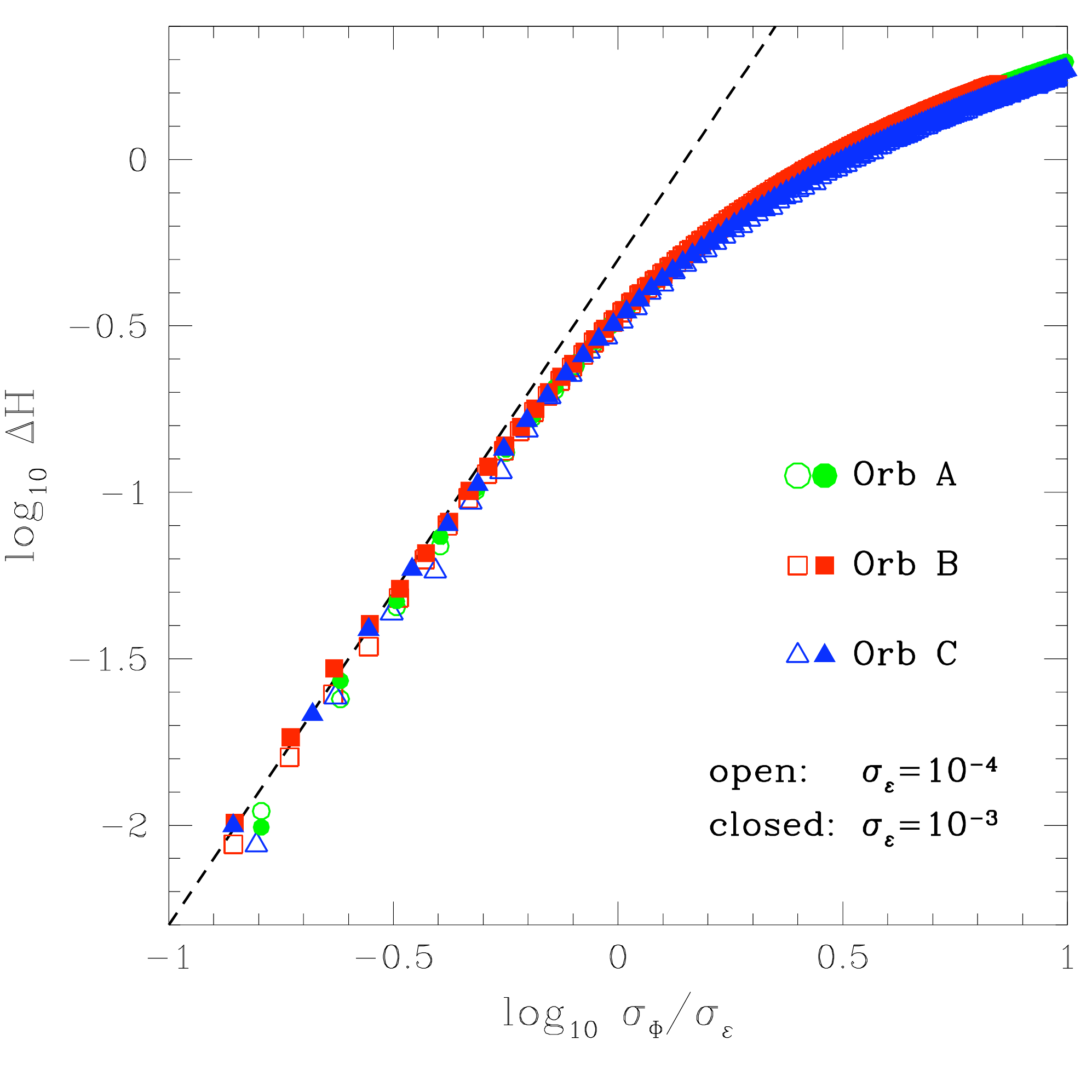}
\caption{ Variation of entropy as a function of the dispersion in the potential values ($\sigma_\Phi$) introduced by biases in $\Phi_0$ (similar curves are obtained from biases in $d_0$). The dashed line shows the theoretical expectation from eq.~(\ref{eq:deltah}). Note that for small biases in the potential the entropy of the orbital energy distribution goes as $\Delta H\simeq 1/2 (\sigma_\Phi/\sigma_\varepsilon)^2$.}
\label{fig:delent}
\end{figure}

Fig.~\ref{fig:delent} shows the relative variation of entropy $\Delta H$ as a function of $\sigma_\Phi/\sigma_\varepsilon$, where $\sigma_\Phi=\sqrt{\langle \delta \Phi^2\rangle - \langle \delta \Phi \rangle ^2}$ is the dispersion in the potential introduced by a model bias. For small potential biases ($\sigma_\Phi\ll \sigma_\varepsilon$) the entropy increase scales as $\Delta H=1/2 (\sigma_\Phi/\sigma_\varepsilon)^2$ independently of orbital eccentricity and energy dispersion, as expected from eq.~(\ref{eq:deltah}). For $\sigma_\Phi\gg \sigma_\varepsilon$ the Taylor expansion used in \S\ref{sec:entropy} does not hold, and deviations between the analytical expectation and the numerical results start to arise. These results provide the basis of our method for measuring the Milky Way potential, as they show that {\it the true Milky Way potential is that which minimizes the entropy of cold tidal debris with separable energy distributions}.

\subsection{Biases arising from the potential form}\label{sec:form}
\begin{figure}
\includegraphics[width=84mm]{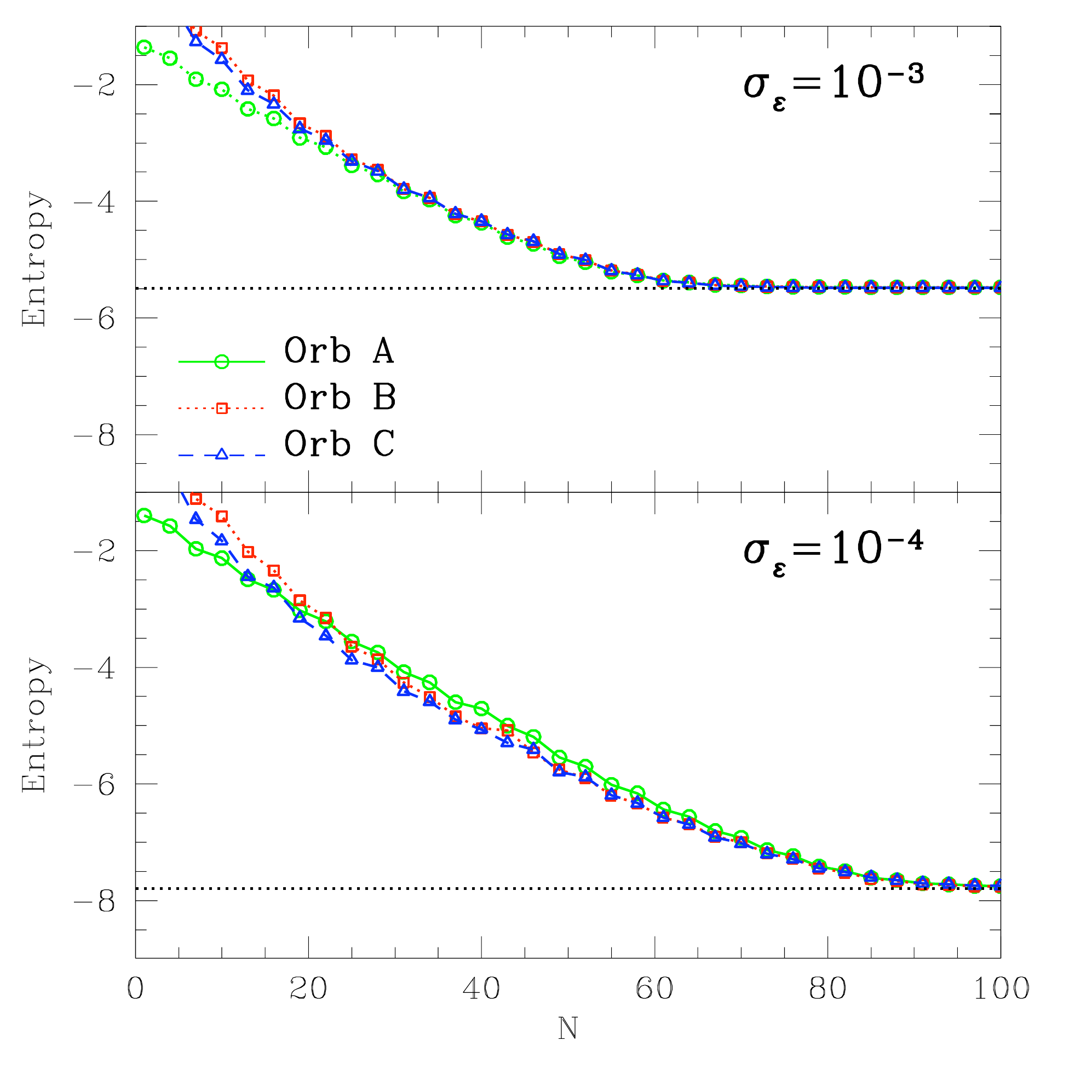}
\caption{Effects of a bias in the functional form of the potential model. Here we expand the host potential model as $\tilde \Phi(r)=2\Phi_0\sum_{k=0}^{(N-1)/2} y^{2k+1}/(2k+1)] + \Phi_0\ln d_0^2$, with $y=(r/d_0)^2/[2+(r/d_0)^2]$; hence the true logarithmic potential of eq.~(\ref{eq:poten}) is recovered in the limit $N\rightarrow \infty$. For ease of reference we plot dotted lines to mark the entropy of a Gaussian distribution. As expected, the entropy asymptotically converges to its minimum as the number of terms in the potential expansion increases.}
\label{fig:logexp}
\end{figure}

Biases may also arise from a wrong parametrization of the Milky Way potential. This is illustrated in Fig.~\ref{fig:logexp} by means of a simple experiment. Let us consider the following potential
\begin{eqnarray}
\tilde \Phi(r) = 2\Phi_0\bigg[y + \frac{y^3}{3} + ... + \displaystyle\sum_{k=0}^{(N-1)/2} \frac{y^{2k+1}}{2k+1}\bigg] + \Phi_0\ln d_0^2, 
\label{eq:logexp}
\end{eqnarray}
where $y=(r/d_0)^2/[2+(r/d_0)^2]$. One can easily show that $\lim_{N\rightarrow \infty}\tilde \Phi = \Phi$, hence we expect a minimum in the entropy in that limit. This is clearly seen in Fig.~\ref{fig:logexp}, which shows an entropy that decreases toward the entropy of the Gaussian distribution (dotted lines) as the number of terms in the potential series increases. Thus, this exercise highlights the possibility of using the entropy of stellar tidal debris not only to find the best parameters of a Galaxy potential, but also to distinguish between different parametrizations.

\subsection{Biases arising from the adopted law of gravity}\label{sec:law}
Let us now consider other gravity theories that allow for the definition of an integral of motion equivalent to the Newtonian orbital energy. From \S\ref{sec:entropy} it follows that {\it choosing} the wrong gravity or cosmology shall introduce a bias in the orbital energy values, which is bound to increase the entropy associated to the energy distribution of tidal debris. 

\subsubsection{Time-variability of $G$}\label{sec:G}
Let us examine first a cosmology where gravity is Newtonian, but $G$ evolves with time. For example under Dirac's large-number hypothesis (Dirac 1938)\footnote{See Uzan (2003) for a compilation of experimental bounds on the variation of $G$ with time.}
\begin{eqnarray}
\frac{G m_p m_e}{e^2}\simeq 10^{-39}\simeq \frac{e^2}{m_e c^3 t};
\label{eq:dirac}
\end{eqnarray}
where $t$ is the time since the big bang, $m_p$ and $m_e$ are the proton and electron masses, and $e$ is the electron charge. In a Universe where the properties of elementary particles remain constant, $G\propto 1/t$.

By inspection of the equations of motions in a Dirac cosmology, Lynden-Bell (1982) found that the orbital energy per unit mass of a particle moving in a potential $\Phi$ associated to a mass distribution $\rho$ is a constant of motion if written as 
\begin{eqnarray}
E_D = H_0^2 t^2\bigg[ \frac{1}{2}\bigg (\frac{d\bf r}{dt}\bigg)^2 + \frac{G}{G_0}\Phi({\bf r}) - \bigg (\frac{d\bf r}{dt}\cdot \frac{\bf r}{t}\bigg)\bigg] + \frac{1}{2}H_0^2{\bf r}^2;
\label{eq:ed}
\end{eqnarray}
where $\nabla^2\Phi=4\pi G_0 \rho$, $H_0$ is the Hubble constant and $G=G_0/(H_0t)$. 
Here the sub-index ``$D$'' denotes quantities calculated in a Universe where $G\propto 1/t$. 

Thus, by analogy with eq.~(\ref{eq:phibias}) the energy bias introduced by the cosmological model at the present time, $t=H_0^{-1}$, is $\delta \Phi_D=\pm [-H_0 ({d\bf r}/dt\cdot {\bf r}/t)+ 1/2H_0^2{\bf r}^2]$, where the minus and the plus symbols correspond to Universes with a constant and an evolving $G$, respectively. Two points are noteworthy: first, $\delta \Phi_D$ only vanishes in the case of centrally-located ($r=0$) orbits; in the case of circular orbits, for which $d{\bf r}/dt\cdot {\bf r}=0$, it reduces to $1/2H_0^2{\bf r}^2$. And second, the energy bias introduced by our choice of gravitational potential (see \S\ref{sec:param} and~\S\ref{sec:form}) is independent of that associated to our choice of cosmology. In practical terms this implies that the Milky Way potential and a possible time-variability of $G$ can be constrained simultaneously by minimizing the entropy of stellar tidal debris. 

To illustrate this point let us re-calculate the entropy of the models introduced in \S\ref{sec:models} using eq.~(\ref{eq:ed}). The presence of $H_0$ introduces a physical scale in the solution, which forces us to scale the test-particle models to physical units. Given that our method will be mostly applied to Milky Way objects, we choose $\Phi_0=1/2 (220 \kms)^2$, $d_0=12$ kpc and $t=H_0^{-1}=14$Gyr. A comparison between the Newtonian and Dirac entropy variation as a function of biases in $\Phi_0$ and $d_0$ are shown in Fig.~\ref{fig:grav}. For this particular example the test-particle models are calculated in a Newtonian potential, so the case $G={\rm const.}$ (solid lines) yields $\Delta H=0$ for $\delta \Phi_0=\delta d_0=0$. Dotted lines show the effects of assuming $G\propto 1/t$. As expected, the entropy is higher than in the Newtonian case independently of the assumed values for $\Phi_0$ and $\d_0$. Note however the effects of a time-variable $G$ are mostly visible at $\delta \Phi_0\approx \delta d_0\approx 0$.

\subsubsection{MOND}\label{sec:mond}
The results obtained in the above Section also apply to gravity theories that are not Newtonian. The modified Newtonian gravity QMOND (Milgrom 2010) is an interesting example. In this theory the gravitational force per unit mass can be simply written as
\begin{eqnarray}
{\bf g}_M={\bf g}_N\nu(r)\equiv{\bf g}_N \bigg(\frac{1}{2} + \sqrt{\frac{1}{4} + \frac{a_0}{g_N}}\bigg),
\label{eq:mondf}
\end{eqnarray}
where $a_0\approx 1.2\times 10^{-10} {\rm m/s}^{2}$ is Milgrom's constant, and $g_N=-G M(<r)/ r^{2}$ is modulus of the Newtonian specific force. In the deep-MOND regime, $g_N\ll a_0$, we have $g_M\approx \sqrt{ a_0 g_N}$; wheres if $g_N\gg a_0$ the Newtonian solution is recovered. In systems with spherical symmetry the corresponding gravitational potential is
\begin{eqnarray}
\Phi_M(r) = \int_r^\infty g_M(r')\d r';
\label{eq:phimond}
\end{eqnarray}
whereas the mass profile associated to the logarithmic potential of eq.~(\ref{eq:poten}) can be written as
\begin{eqnarray}
M(<r) = \frac{2\Phi_0}{G}\frac{r^3}{r^2+d_0^2}.
\label{eq:massmond}
\end{eqnarray}
Clearly, the Newtonian acceleration reaches a maximum at $r=d_0$. For the fiducial parameters $\Phi_0=1/2 (220 \kms)^2$ and $d_0=12$ kpc, we have that $g_N/a_0(r=d_0)= \Phi_0/(d_0a_0) \simeq 0.544$; hence the minimum Mondian-to-Newtonian ratio is $\nu(r=d_0)\simeq 1.944$. 

\begin{figure}
\includegraphics[width=84mm]{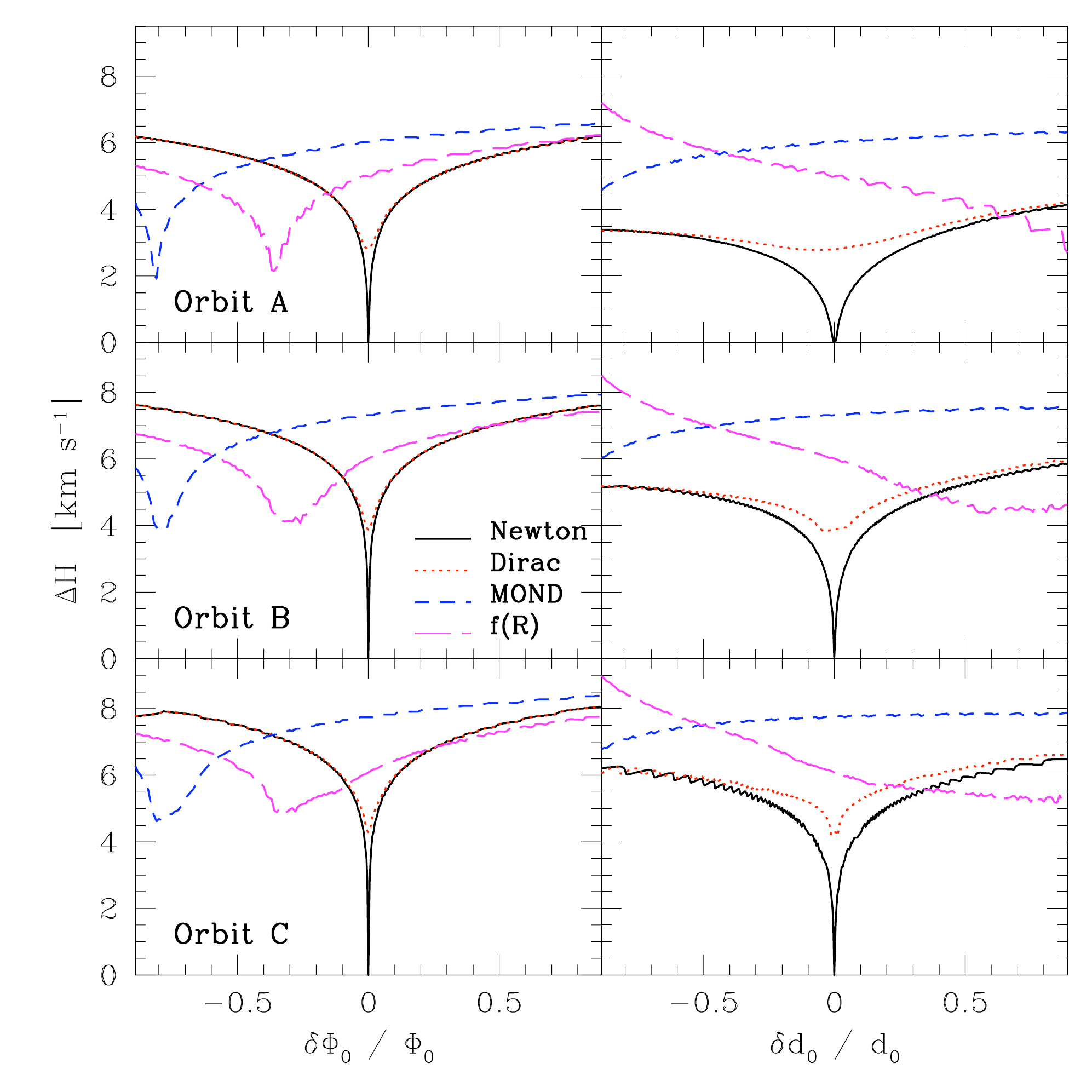}
\caption{Increase in entropy due to biases in the potential parameters $\Phi_0$, $d_0$, as well as in the gravity theory, for a model with an unbiased energy dispersion $\sigma_\ve=10^{-4} \Phi_0$. Orbits are integrated in a Newtonian potential (solid lines). As expected, the entropy finds a minimum ($\Delta H=0$) when the biases in the potential parameters vanish, i.e. $\delta \Phi_0=\delta d_0=0$. To allow a comparison between different gravity theories we scale these parameters to $\Phi_0=(220\kms)^2/2$ and $d_0=12$ kpc, so that orbital energies are measured in units of $(\kms)^2$. Dotted and dashed lines show the effects of assuming Dirac's cosmology (i.e. $G\propto 1/t$) and a MONDian gravity, respectively. Long-dashed lines show the same calculation in a $f(R)$ gravity theory with $\beta=0.82$ ($n\approx 3.5)$ and $r_c=d_0$ (see \S\ref{sec:fr}).  }
\label{fig:grav}
\end{figure}

The differences between $\Phi_M$ and $\Phi_N$ are thus much stronger than those introduced by a time-varying $G$ (\S\ref{sec:G}). A comparison between the dashed and solid lines in Fig.~\ref{fig:grav} shows that $\Delta H$ is dominated by the modification of the Newtonian gravity given by eq.~(\ref{eq:mondf}), rather than by the biases in the potential parameters. It is interesting to notice that the minimum entropy occurs for $\delta \Phi_0\simeq -0.8 <0$, as one would expect given that the MOND potential is stronger than the Newtonian one at all radii.

\subsubsection{$f(R)$ gravities}\label{sec:fr}
Another interesting case of gravity models that aim to modify Einstein's General Relativity at large scales can be found in $f(R)$ theories. In these models the action can be written as
\begin{eqnarray}
\mathcal{A}=\int \d^4x\sqrt{-g}[f(R) + \mathcal{L}_m];
\end{eqnarray}
here $f(R)$ is a generic function of the Ricci scalar curvature that is usually assumed to follow a power-law function, i.e. $f(R)=f_0R^n$; $g_{\mu\nu}$ is the metric, and $\mathcal{L}_m$ is the standard matter Lagrangian. Einstein's General Relativity with contribution of a cosmological constant is recovered for $n=1$ and $f(R)=R+2\Lambda$.

Capozziello et al. (2007) explore modifications of the Newtonian gravity where the potential of a spherical shell of mass $\d m$ can be written as
\begin{eqnarray}
\d\Phi = -\frac{G \d m}{2 r}\bigg[1 + \bigg(\frac{r}{r_c}\bigg)^\beta\bigg];
\end{eqnarray}
where
$$\beta\approx \frac{12n^2 - 7n -1 -\sqrt{36 n^4 +12 n^3 -83 n^2 +50 n +1}}{6n^2 -4n +2};$$ so that the relativistic case $n=1$ corresponds to $\beta=0$. Cases where $1- \beta> 0$ yield an increase in the gravitational force on scales $r\simgreat r_c$.

The potential of an extended distribution of mass with a density profile $\rho(r)$ can be thus written as $\Phi_R=1/2(\Phi_N + \Phi_c)$, where $\Phi_N$ is the Newtonian potential and 
\begin{eqnarray}
\label{eq:phifr}
\Phi_c(r)= -4\pi G\bigg[\frac{1}{r}\int_0^r \d r' \rho(r')r'^2\bigg(\frac{r}{r_c}\bigg)^\beta + \\ \nonumber
\int_r^\infty \d r' \rho(r')r'\bigg(\frac{r}{r_c}\bigg)^\beta\bigg].
\end{eqnarray}

In contrast to $\beta$, the quantity $r_c$ is not universal and has to be fitted on individual galaxies. Interestingly, Capozziello et al. (2007) find that $n=3.5$ ($\beta\simeq 0.82$) provides a good fit to the rotation curves of several low-surface-brightness galaxies as well as to Hubble diagram derived from Type Ia supernova with no dark matter.

In order to illustrate how entropy may help to constrain the value of $\beta$ from Milky Way phase-space surveys we plot with long-dashed lines in Fig.~\ref{fig:grav} the variation of entropy as a function of biases in $\Phi_0$ and $d_0$ for $\beta=0.82$ and $r_c=d_0$. Here we consider test-particle models evolved in a Newtonian gravity with an unbiased energy dispersion $\sigma_\ve=10^{-4}\Phi_0$. Models are scaled to the following physical units, $\Phi_0=(220\kms)^2/2$ and $d_0=12$ kpc, and energies are measured in units of $(\kms)^2$.
As with the MOND theory, the fact that the modified gravity is stronger than the Newtonian one at all radii leads to an entropy that is minimum at $\delta \Phi_0\simeq -0.4 <0$, and to a scale $d_0$ that is poorly constrained.

As a final remark we note that 
similar restricted N-body experiments can be carried in Dirac, MONDian and $f(R)$ gravities, where for a given mass distribution the minimum entropy associated to the true gravity theory must be lower than in the Newtonian case independently of the potential parametrization.

\begin{figure*}
\includegraphics[width=170mm, height=170mm]{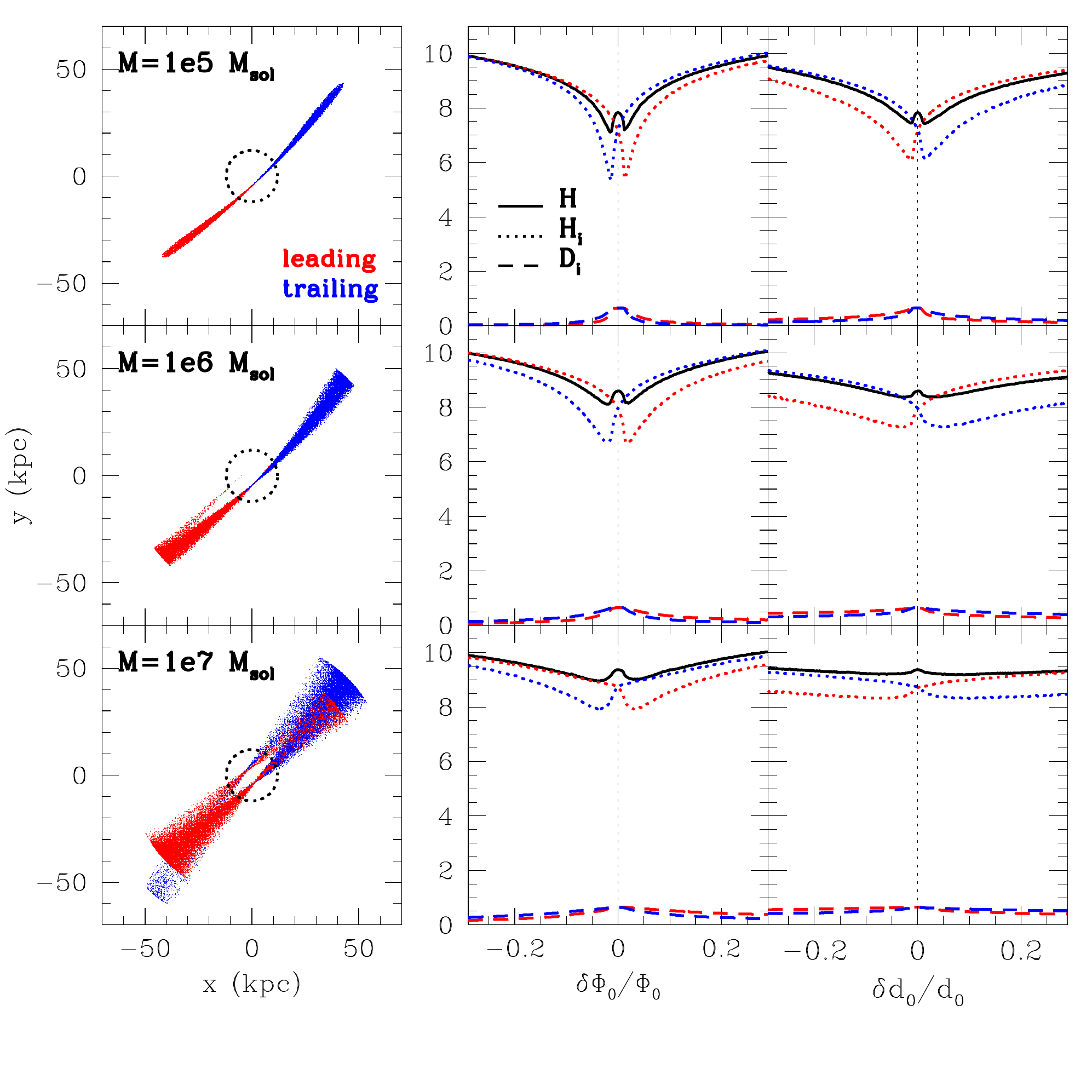}
\caption{{\bf Left panel}: Spatial distribution of tidal debris. Particles are coloured-coded depending on whether they belong to the leading (in red) and trailing (in blue) tails. Dotted lines mark the radius $r=d_0$. {\bf Middle} and {\bf right panels}: Variation of the debris entropy ($H$, solid lines) as a function of biases in the values of $\Phi_0$ and $d_0$, respectively. Energies are measured in units of $(\kms)^2$. The entropy ($H_i$) and the KL-divergence ($D_i$) of the individual tails are shown with dotted and dashed lines, respectively. As expected from eq.~(\ref{eq:deltah}) the entropy of the individual tails increases under the presence of biases in the calculus of the host potential. Note that the opposite trend is visible for the variation of $D_i$, which is maximum for the unbiased $\delta \Phi_0=\delta d_0=0$ potential. }
\label{fig:KL}
\end{figure*}
\section{Effects of Self-gravity}\label{sec:self}
Tidal streams formed in realistic galaxy potentials are poorly represented by single orbits (e.g. Pe\~narrubia et al. 2006; Eyre \& Binney 2011). Numerical experiments show that unbound particles escape through the Lagrange L1 and L2 points of the disrupting object (e.g. Renaud et al. 2011), forming leading and trailing tails with orbital energies that are respectively higher and lower than that of the progenitor system. Given that tidal tails have overlapping energy distributions (Pe\~narrubia et al. 2006) and occupy different volumes in phase-space, it follows from \S\ref{sec:dh_mult} that the progenitor's self-gravity must introduce a bias in our method. 

To determine the extent to which the leading and trailing tails have different energy distributions it is useful to measure the Kullback-Leibler (1951) divergence of the individual tails, which is defined as
\begin{eqnarray}
D_i = \int f_i(\ve) \ln\bigg[\frac{f_i(\ve)}{f(\ve)}\bigg] \d\ve \equiv -H_i + H_{c,i}
\label{eq:di}
\end{eqnarray}
where the sub-index $i=l,t$ denotes leading and trailing tail distributions, respectively. The right-hand term $H_{c,i}=-\int f_i(\ve) \ln f(\ve) \d\ve$ is called {\it crossed entropy}. 

In the case of a separable energy distribution one has that $f_i(\ve)=f(\ve)$, so that the crossed entropy of the tails is equal to the overall entropy, i.e $H_{c,i}=H$; hence, the KL-divergence is $D_i=0$. The Kullback-Leibler (or KL) divergence provides therefore a measure of the separability of the orbital energy distribution.

Using the notation of \S\ref{sec:dh_mult} we have that $f=\alpha f_l + (1-\alpha)f_t$, so that the overall entropy can be written as 
\begin{eqnarray}
H=-\int f(\ve)\ln f(\ve) \d\ve = \\ \nonumber
-\alpha\int f_l(\ve)\ln f(\ve)\d\ve - (1-\alpha)\int f_t(\ve)\ln f(\ve)\d\ve\equiv\\ \nonumber
\alpha H_l + (1-\alpha)H_t + \alpha D_l  + (1-\alpha)D_t\equiv \langle H\rangle_{l,t} + \langle D \rangle_{l,t};
\label{eq:di}
\end{eqnarray}
where $\langle H\rangle_{l,t}$ and $\langle D\rangle_{l,t}$ are the combined entropy and KL-divergence of the tidal tails.

To illustrate how the  KL-divergence varies as a function of biases in the calculus of orbital energy we run self-consistent N-body simulations of tidally-disrupting clusters moving on the orbit C in the potential defined by eq.~(\ref{eq:poten}). For ease of reference the potential parameters are set to $\Phi_0=(220\kms)^2/2$ and $d_0=12$kpc. The cluster models follow a cored Dehnen (1993) profile, $\rho_c = (3M/4\pi)a/(a+r)^{4}$, with masses $M=10^5, 10^6$ and $10^7M_\odot$ and radii $a=0.29, 0.62$ and 1.33 kpc, respectively, which are therefore comparable to the dynamical masses and sizes of dwarf spheroidal galaxies (e.g. Walker et al. 2009). The tidal evolution of the clusters is followed using {\sc superbox} (see Fellhauer et al. 2000 for details) for 7.35 Gyr. The numerical resolution and time-steps are chosen so that over that time interval the energy is conserved at a level $\Delta \ve \simless 0.01 G M/a$ for models evolved in isolation.

The left panel of Fig.~\ref{fig:KL} shows the spatial distribution of debris projected on the progenitor's orbital plane. Here particles are coloured-coded according to their orbital energies: red (leading) and blue (trailing) particles denote energies that are lower and higher, respectively, than that of the progenitor system. In all models the number of particles in the leading and trailing tails is practically the same, so $\alpha\approx 1/2$.
Note that for a fixed orbit and integration time it takes many more orbits for cold streams to spread a substantial fraction of $2\pi$ in orbital phase. 

The middle and right panels show the variation in entropy and KL-divergence as a function of biases in $\Phi_0$ and $d_0$. A few interesting points are worth noticing. First, in contrast to the tails entropy ($H_i$) the KL-divergence ($D_i$) finds a {\it maximum} for the unbiased potential $\delta \Phi=0$. For energies measured in units of $(\kms)^2$ we find that the maximum combined KL-divergence is $\langle D\rangle_{l,t}\simeq 0.65$, a value that barely depends on the progenitor mass. In contrast the combined entropy of the tails in the unbiased potential is $\langle H \rangle_{l,t}\simeq 7.17, 7.95$ and 8.72 for $M=10^5, 10^6$ and $10^7M_\odot$, respectively.
As expected, the ratio between the combined KL-divergence and the entropy of the tails ($\langle D\rangle_{l,t}/\langle H\rangle_{l,t}$) is small, which indicates that tidal debris of low-mass objects follow nearly-separable energy distributions. However, it is worth noting that although this quantity becomes smaller as the progenitor mass decreases, it only vanishes in the limit $M\rightarrow 0$.

Note also that if each of the tails followed a separable energy distribution, then from \S\ref{sec:dh_single} one should expect $\Delta H_i\simeq (\sigma_{\Phi,i}/\sigma_{\ve,i})^2$ in the limit $\delta \Phi/\Phi\ll 1$. However, this is clearly not the case. The reason can be found in the time-variation of the progenitor's self-gravity. As mass stripping progresses the location of the Lagrange points L1 and L2 move closer to the centre of the disrupting system, which introduces a dependence between the energy of the particles and the time when they become unbound. This inevitable effect is responsible for the asymmetric (and mirrored) dependence of $H_l$ and $H_t$ as a function of $\delta \Phi_0$ and $\delta d_0$, which is not visible in Fig.~\ref{fig:ent_4}.

Finally, the fact that $\langle D \rangle_{l,t}\neq 0$ as well as a decreasing function of the energy bias has two undesirable effects. First, $H$ is less sensitive to the presence of biases in the energy calculation than $\langle H\rangle_{l,t}$. Second, the condition for minimum entropy $H'=\langle H \rangle'_{l,t} + \langle D \rangle'_{l,t}=0$, where $f'\equiv \d f/\d\delta\Phi$, 
is now met for $\langle H \rangle'_{l,t} = \langle D \rangle_{l,t}'=0$, which yields a local maximum at $\delta \Phi\approx 0$; and for $|\langle H \rangle'_{l,t}| = -|\langle D \rangle'_{l,t}|$, which yields two local minima around $\delta \Phi\approx 0$. The progenitor's self-gravity therefore introduces a bias in the Minimum Entropy Method. For the N-body satellite models explored here, which have dynamical masses $M\simless 10^7M_\odot$ and follow eccentric orbits about a Milky Way-like potential, the bias is fairly small, $|\delta \Phi|\simless 0.03\Phi_0$. 
However, notice that the bias does not only depend on the progenitor's self-gravity. The fact that $\Delta H_i \simeq (\sigma_{\Phi,i}/\sigma_{\ve,i})^2$ at $\delta \Phi\approx 0$ implies that the steepness of the function $\langle H \rangle'_{l,t}$, and thus the location of the two minima, will also depend on the orbit of the progenitor about the host potential that is being inspected. 

These results suggest that the effects introduced by the progenitor's self-gravity can be lessened in two ways. First, one may attempt to minimize the ratio $\langle D\rangle_{l,t}/|\langle H\rangle_{l,t}|$ by applying the Minimum Entropy Method to tidal debris that originate from low-mass systems. In addition, one may derive the probability that a given star belongs either to the leading or trailing tail by inspection of the distribution of debris in the integral-of-motion space. 
As shown in Fig.~\ref{fig:KL}, knowing the membership probability allows a derivation of the crossed-entropy of the individual tails, giving further sensitivity to biases in the host potential.

\section{Summary and discussion}\label{sec:disc}
In this contribution we have introduced the {\it Minimum Entropy Method}. This statistical technique is devised to constrain the Milky Way gravitational potential and test different gravity theories directly from stellar 6D phase-space catalogues and without adopting dynamical models. Our method rests upon two fundamental assumptions: (1) the gravity theory under study allows for the existence of an orbital energy that is an integral of motion; and (2) cold structures in phase-space resulting from the tidal disruption of gravitationally-bound systems have an orbital energy distribution that is separable in energy and space. Then it is shown in \S\ref{sec:entropy} that any bias in the calculus of the orbital energy (due to the adoption of an incorrect potential and/or gravity theory) translates into an increase of the debris entropy. 
Examining what type of gravity theories obey the first condition goes beyond our current goals, but it is worth considering here in what cases the second condition may not be met. 

In \S\ref{sec:self} we show that the progenitor's self-gravity induces the formation of leading and trailing tidal tails, which occupy different phase-space volumes and follow distinct energy distributions. Because the separability condition is broken, self-gravity introduces a bias in the location of the minimum entropy, thus limiting the application of the method to tidal debris that originate from low-mass systems. However, the effects of self-gravity can be lessened by deriving the probability that a star in the debris sample belongs either to the leading or trailing tail. With this information at hand it is possible to quantify systematic variations of the energy distribution in terms of the cross-entropy of individual tails and subtract the contribution of the non-separable term $\langle D\rangle_{l,t}$ (the so-called KL-divergence) from the overall entropy $H$. The resulting quantity is the combined entropy of the leading and trailing tails $\langle H\rangle_{l,t} = H - \langle D\rangle_{l,t}$, which is more sensitive to biases in the host gravity than the overall entropy.

Perturbations on the orbit of a system undergoing tidal disruption may also break the separability of the energy distribution. For example, a drag force (e.g. dynamical friction) acting upon a tidally-disrupting system introduces a dependence between the time when stars are lost to tides and their mean orbital energy. Similarly, interactions with bound substructures lingering in the Milky Way halo (e.g. molecular clouds, stellar clusters) are bound to introduce similar biases in our method. But given that the rate of two-body encounters scales with the progenitor mass (see e.g. \S7.5 of Binney \& Tremaine 2008) both effects can be again minimized by applying the Minimum Entropy Method to tidal debris that originate from low-mass systems.

Finally, it is well known that observational errors plus the presence of a background in the debris sample may also introduce biases in methods of statistical inference. For example, on account of our position within the Galaxy a systematic increase in the position and velocity errors with heliocentric distance seems unavoidable. The large number of ways in which errors and systematic biases may arise in phase-space catalogues prevents a generic approach to this problem, being more convenient to inspect this issue in individual cases.

We plan to examine these and other aspects of our method in a forthcoming contribution, where realistic simulations of self-gravitating stellar clusters undergoing tidal disruption will be built by means of N-body techniques. Also, it will prove an exciting exercise to explore to what extent local gravity tests may complement those devised on cosmological scales (e.g. Zhao, Peacock \& Li 2012) with help of mock Gaia catalogues.  

\vskip1cm
The authors thank Hong-Sheng Zhao for his helpful comments on QMOND. JP wishes to thank Mike Irwin and Lu\'is Aguilar for their useful insights on entropy. JP acknowledges support from the Ram\'on y Cajal Program as well as by the Spanish grant AYA2010-17631 awarded by the Ministerio of Econom\'ia y Competitividad. MGW is supported by NASA through Hubble Fellowship grant HST-HF-51283.01-A, awarded by the Space Telescope Science Institute, which is operated by the Association of Universities for Research in Astronomy, Inc., for NASA, under contract NAS5-26555.

{}

\end{document}